\documentclass[twocolumn,aps,prb,showpacs]{revtex4}
\usepackage[T1]{fontenc}
\usepackage[latin9]{inputenc}
\usepackage{mathrsfs}
\usepackage{amsmath}
\usepackage{amssymb}
\usepackage{graphicx}
\usepackage{esint}

\makeatletter

\providecommand{\tabularnewline}{\\}

\@ifundefined{textcolor}{}
{%
 \definecolor{BLACK}{gray}{0}
 \definecolor{WHITE}{gray}{1}
 \definecolor{RED}{rgb}{1,0,0}
 \definecolor{GREEN}{rgb}{0,1,0}
 \definecolor{BLUE}{rgb}{0,0,1}
 \definecolor{CYAN}{cmyk}{1,0,0,0}
 \definecolor{MAGENTA}{cmyk}{0,1,0,0}
 \definecolor{YELLOW}{cmyk}{0,0,1,0}
}

\makeatother

\begin{document}

\title{Phase diagram of a non-Abelian Aubry-André-Harper model with $p$-wave
superfluidity}

\author{Jun Wang$^{1}$, Xia-Ji Liu$^{2}$, Gao Xianlong$^{1}$, and Hui
Hu$^{1,2}$}

\email{hhu@swin.edu.au}

\affiliation{$^{1}$Department of Physics, Zhejiang Normal University, Jinhua
321004, China}

\affiliation{$^{2}$Centre for Quantum and Optical Science, Swinburne University
of Technology, Melbourne 3122, Australia}

\date{\today}
\begin{abstract}
We theoretically study a one-dimensional quasi-periodic Fermi system
with topological $p$-wave superfluidity, which can be deduced from
a topologically non-trivial tight-binding model on the square lattice
in a uniform magnetic field and subject to a non-Abelian gauge field.
The system may be regarded a non-Abelian generalization of the well-known
Aubry-André-Harper model. We investigate its phase diagram as functions
of the strength of the quasi-disorder and the amplitude of the $p$-wave
order parameter, through a number of numerical investigations, including
a multifractal analysis. There are four distinct phases separated
by three critical lines, i.e., two phases with all extended wave-functions
(I and IV), a topologically trivial phase (II) with all localized
wave-functions and a critical phase (III) with all multifractal wave-functions.
The phase I is related to the phase IV by duality. It also seems to
be related to the phase II by duality. Our proposed phase diagram
may be observable in current cold-atom experiments, in view of simulating
non-Abelian gauge fields and topological insulators/superfluids with
ultracold atoms. 
\end{abstract}


\maketitle

\section{Introduction}

The Aubry-André-Harper (AAH) model is a workhorse for the study of
dynamics of particles in one-dimensional (1D) quasi-periodic systems
\cite{Harper1955,Aubry1980}. Over the past few decades, it has been
extensively used to theoretically understand the transport and Anderson
localization properties of these interesting systems, revealing a
variety of transitions between metallic (extended), critical and insulator
(localized) phases \cite{Hiramoto1992,Kohmoto1983a,Ostlund1983,Kohmoto1983b,Thouless1983,Hiramoto1989,Geisel1991,Han1994,Chang1997,Takada2004,Liu2015}.
Most recently, the AAH model has attracted renewed attentions due
to its experimental realization in photonic crystals \cite{DalNegro2003,Lahini2009,Kraus2012}
and ultracold atoms \cite{Roati2008,Modugno2010}. It has been found
to play a non-trivial role to characterize emerging topological states
of matter \cite{Lang2012a,Lang2012b,DeGottardi2013,Cai2013,Ganeshan2013,Satija2013,Kraus2013,Zhu2013,Grusdt2013,Barnett2013,Deng2014}
and the intriguing phenomenon of quantum many-body localization \cite{Iyer2013,Schreiber2015}.

The AAH model can be formally derived from the reduction of a two-dimensional
(2D) quantum Hall system to a 1D chain \cite{Hofstadter1976}. By
taking a Landau gauge for the magnetic field with $\phi$ flux quanta
per unit cell, the motion of an electron in a 2D rectangular lattice
with tight-binding hopping strengths $t_{x}$ and $t_{y}$ can be
described by the Hofstadter Hamiltonian ($\theta_{ij}^{y}=2\pi i\phi$)
\cite{Hofstadter1976},

\begin{equation}
\mathcal{H}_{2D}=\sum_{ij}\left[\hat{a}_{i+1,j}^{\dagger}t_{x}\hat{a}_{i,j}+e^{i\theta_{ij}^{y}}\hat{a}_{i,j+1}^{\dagger}t_{y}\hat{a}_{i,j}+\textrm{H.c.}\right],\label{eq:HofstadterHamiltonian}
\end{equation}
which has a translational symmetry in the $y$-direction. As a result,
the momentum of the motion along the $y$-axis, $k_{y}\subseteq[0,2\pi)$,
is well defined. Defining a reduced field operator $\hat{c}_{i}$
via $\hat{a}_{i,j}=e^{-ik_{y}j}\hat{c}_{i}/\sqrt{L_{y}}$, where $L_{y}$
is the number of sites along the $y$-axis, we deduce from $\mathcal{H}_{2D}$
the AAH model,

\begin{equation}
\mathcal{H}=\sum_{i}\left[t_{x}\left(\hat{c}_{i+1}^{\dagger}\hat{c}_{i}+\textrm{H.c.}\right)+t_{y}V_{i}\hat{c}_{i}^{\dagger}\hat{c}_{i}\right],
\end{equation}
where $V_{i}\equiv2\cos(2\pi i\phi+k_{y})$. The localization properties
of the AAH model can be easily understood from the parent Hofstadter
Hamiltonian. When $t_{y}>t_{x}$, the electron prefers to hop along
the $y$-direction and its wave-function thus becomes localized in
the $x$-axis. In contrast, when $t_{y}<t_{x}$, the electron motion
will extend over the entire $x$-axis. Indeed, these two cases are
related by the so-called Aubry-André duality, which can be easily
derived by considering two different realizations of gauge for the
magnetic field \cite{Aubry1980}. On the other hand, the topological
properties of the AAH model can also be understood from the 2D Hofstadter
Hamiltonian that underlies the topologically non-trivial quantum Hall
phenomenon.

In this work, motivated by the recent proposals of simulating non-Abelian
gauge fields \cite{Osterloh2005,Goldman2009,Goldman2014} and topological
insulators/superfluids \cite{Liu2012a,Liu2012b} with ultracold atoms,
we consider a generalized Hofstadter Hamiltonian, obtained by using
a $N$-component field operator $\mathbf{\hat{a}}_{i,j}=[\hat{a}_{i,j}^{(1)},\cdots,\hat{a}_{i,j}^{(N)}]^{T}$
in Eq. (\ref{eq:HofstadterHamiltonian}) and by replacing the hopping
strengths $t_{x}$ and $t_{y}$ with two SU($N$) matrices $\hat{T}_{x}$
and $\hat{T}_{y}$. Taking the same reduction to a 1D chain along
the $x$-direction (i.e., using $\hat{a}_{i,j}^{(p)}=e^{-ik_{y}j}\hat{c}_{i}^{(p)}/\sqrt{L_{y}}$
with $p=1,\cdots,N$), we then obtain a generalized non-Abelian AAH
model,

\begin{equation}
\mathcal{H}=\sum_{i}\left[\left(\hat{\mathbf{c}}_{i+1}^{\dagger}\hat{T}_{x}\hat{\mathbf{c}}_{i}+\textrm{H.c.}\right)+V_{i}\hat{\mathbf{c}}_{i}^{\dagger}\hat{T}_{y}\hat{\mathbf{c}}_{i}\right],\label{eq:nonAbelianHarperModel}
\end{equation}
where for simplicity we have assumed that $\hat{T}_{y}^{\dagger}=\hat{T}_{y}$
and have used $\hat{\mathbf{c}}_{i}=[\hat{c}_{i}^{(1)},\cdots,\hat{c}_{i}^{(N)}]^{T}$.

To be concrete, we consider the simplest non-Abelian case with two-component
field operator and the SU($N=2$) hopping matrices 
\begin{equation}
\hat{T}_{x}=t_{x}\hat{\sigma}_{z}-i\Delta\hat{\sigma}_{y}\label{eq:Tx}
\end{equation}
and 
\begin{equation}
\hat{T}_{y}=t_{y}\hat{\sigma}_{z},\label{eq:Ty}
\end{equation}
where $\hat{\sigma}_{y}$ and $\hat{\sigma}_{z}$ are the usual 2
by 2 Pauli matrices. In the absence of the quasi-periodic disorder
term (i.e, $t_{y}=0$), the model describes a 1D topologically non-trivial
insulator or a spinless $p$-wave superfluid \cite{Hasan2010,Qi2011,Shen2012}.
On the other hand, in the limit of $\Delta=0$, where $\hat{T}_{x}$
and $\hat{T}_{y}$ commute with each other, the model reduces to the
standard AAH model for each component.

\begin{figure}
\begin{centering}
\includegraphics[clip,width=0.48\textwidth]{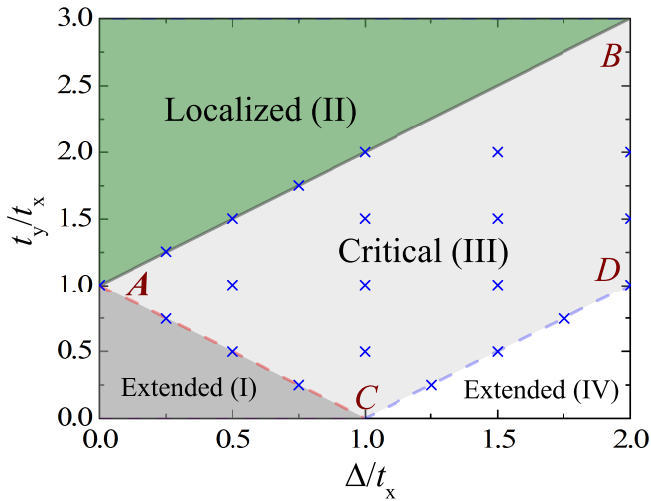} 
\par\end{centering}

\protect\caption{(Color online). Phase diagram of a quasi-disordered Fermi system with
a \textit{p}-wave order parameter $\Delta$ and disorder strength
$t_{y}$. The energy is in units of the hopping amplitude $t_{x}$.
Four different phases are separated by the critical lines $AB$, $AC$,
and $CD$: the phase (I) with spatially extended wave-functions, the
topologically trivial phase (II) with localized wave-functions, the
critical phase (III) with fractal wave-functions, and another extended
phase (IV). The phases (II) and (IV) are dual to the phase (I). Multifractal
analysis of wave-functions has been made for the 21 points in the
critical phase (III) or on the critical lines (see Table I).}

\label{fig1} 
\end{figure}

We find that the localization and topological properties of the AAH
model are profoundly affected by the existence of a nonzero $\Delta$.
Our main results are summarized in the phase diagram Fig. \ref{fig1}.
There are four distinct phases (I-IV) separated by three critical
lines $AB$, $AC$ and $CD$. For a sufficiently large quasi-disorder
strength, i.e., above the line $AB$ where $t_{y}>t_{x}+\Delta$ the
system becomes localized and topologically trivial \cite{DeGottardi2013,Cai2013}.
Below the line $AC$ or its duality line $CD$, where $t_{y}<\left|t_{x}-\Delta\right|$,
all the wave-functions of the system are extended. The area enclosed
by the three separation lines is critical and all the wave-functions
are multifractal. In this study, the phase diagram Fig. \ref{fig1}
is theoretically analyzed by a number of numerical approaches, including
a multifractal analysis for the critical area.

The rest of the paper is organized as follows. In the next section
(Sec. II), we explicitly write down the Schrödinger equation of our
1D quasi-periodic system and highlight its particle-hole symmetry
and duality. In Sec. III, we determine the phase diagram by calculating
the inverse participation ratio at a given level of rational approximations
and verifying single-particle wave-functions in different phases.
In Sec. IV, we introduce the multifractal approach and present a scaling
analysis of the structure of wave-functions in the critical regime.
In Sec. V, we discuss the behavior of edge modes or Majorana fermions
in different states. Section VI is devoted to conclusions and outlooks.

\section{Non-Abelian SU(2) Aubry-André-Harper Model}

For the non-Abelian AAH model with the hopping matrices in Eqs. (\ref{eq:Tx})
and (\ref{eq:Ty}), we may write the wave-function 
\begin{equation}
\left|\psi\right\rangle =\sum_{i}\left[u_{i}\hat{c}_{i}^{(1)\dagger}+v_{i}\hat{c}_{i}^{(2)\dagger}\right]\left|0\right\rangle ,
\end{equation}
then the Schrödinger equation $\mathcal{H}\left|\psi\right\rangle =\epsilon\left|\psi\right\rangle $
has the explicit form, 
\begin{eqnarray}
t_{x}\left(u_{i+1}+u_{i-1}\right)+t_{y}V_{i}u_{i}-\Delta\left(v_{i+1}-v_{i-1}\right) & = & \epsilon u_{i},\label{eq:SEa}\\
\Delta\left(u_{i+1}-u_{i-1}\right)-t_{x}\left(v_{i+1}+v_{i-1}\right)-t_{y}V_{i}v_{i} & = & \epsilon v_{i}.\label{eq:SEb}
\end{eqnarray}
Recall that $V_{i}=2\cos(2\pi i\phi+k_{y})$. Following the standard
routine for investigating localization properties of the AAH model,
throughout the paper we set $k_{y}=0$ and take an irrational value
$\phi=(\sqrt{5}-1)/2$. This inverse of the golden mean may be approached
by using the Fibonacci numbers $F_{n}$: $\phi=\lim_{n\rightarrow\infty}F_{n-1}/F_{n},$
where $F_{n}$ is recursively defined by the relation $F_{n+1}=F_{n}+F_{n-1}$,
with $F_{0}=F_{1}=1$. Thus, in numerical calculations we take the
rational approximation 
\begin{equation}
\phi\simeq\phi_{n}=\frac{F_{n-1}}{F_{n}}.
\end{equation}
To minimize the effect of the periodic boundary condition, we assume
that the length of the system is periodic with period $F_{n}$.

At the $n$-th rational approximation, the Schrödinger equation then
becomes periodic with period $F_{n}$. According to Bloch's theorem,
the wave-function $\psi$ can be characterized by a crystal momentum
$q_{x}\subseteq[-\pi/F_{n},+\pi/F_{n})$ and satisfies $\psi_{i+F_{n}}=e^{iq_{x}F_{n}}\psi_{i}$
($i=0,1,\cdots,F_{n}-1$). Therefore, if we represent $\psi$ as a
vector, 
\begin{equation}
\psi=\left[u_{0},v_{0},u_{1},v_{1},\cdots,u_{F_{n}-1},v_{F_{n}-1}\right]^{T},
\end{equation}
the Schrödinger equation can be solved by finding the eigenvalues
and eigenvectors of a $2F_{n}\times2F_{n}$ matrix: 
\begin{equation}
\mathcal{H}_{n}=\left(\begin{array}{ccccccc}
A_{0} & B & 0 & \cdots &  & 0 & C\\
B^{\dagger} & A_{1} & B & 0 & \cdots &  & 0\\
0 & B^{\dagger} & A_{2} & B & 0 & \cdots & 0\\
\vdots & \ddots & \ddots & \ddots & \ddots & \ddots & \vdots\\
0 & \cdots & 0 & B^{\dagger} & A_{F_{n}-3} & B & 0\\
0 &  & \cdots & 0 & B^{\dagger} & A_{F_{n}-2} & B\\
C^{\dagger} &  &  & \cdots & 0 & B^{\dagger} & A_{F_{n}-1}
\end{array}\right),\label{eq:matrixBdG}
\end{equation}
where 
\begin{equation}
A_{i}=2t_{y}\cos\left(2\pi i\phi_{n}\right)\left(\begin{array}{cc}
1 & 0\\
0 & -1
\end{array}\right),
\end{equation}
\begin{equation}
B=\left(\begin{array}{cc}
t_{x} & -\Delta\\
\Delta & -t_{x}
\end{array}\right),
\end{equation}
and 
\begin{equation}
C=\left(\begin{array}{cc}
t_{x} & \Delta\\
-\Delta & -t_{x}
\end{array}\right)\exp\left(-iq_{x}F_{n}\right).\label{eq:matrixBC}
\end{equation}
The energy spectrum consists of $2F_{n}$ bands, each of which is
a function of the crystal momentum $q{}_{x}$. Without loss of generality,
we consider only one state in each band by taking $q_{x}=0$. Numerically,
we solve the eigenvalue and eigenvector problem of the matrix Eq.
(\ref{eq:matrixBdG}) at a given value of $n$ and then perform a
scaling analysis with increasing $n$. The irrational limit is reached
when we extrapolate our numerical results to the scaling limit $1/n\rightarrow0$.

\subsection{Particle-hole symmetry}

With our specific choice of SU(2) hopping matrices, the non-Abelian
AAH model has an interesting particle-hole symmetry. That is, the
model is invariant under the particle-hole transformation: 
\begin{equation}
\hat{c}_{i}^{\left(1\right)}\leftrightarrow\hat{c}_{i}^{\left(2\right)\dagger}.
\end{equation}
In other words, for every particle-like solution $(u_{i},v_{i})$
of the Schrödinger Eqs. (\ref{eq:SEa}) and (\ref{eq:SEb}) with energy
$E\geq0$, there is always a hole-like solution $(v_{i}^{*},u_{i}^{*})$
with energy $-E$. To highlight this particle-hole symmetry, it is
useful to interpret the two components of the system as the particle
and hole components of a spinless $p$-wave Fermi superfluid, in which
the parameter $\Delta$ can be conveniently identified as a $p$-wave
order parameter. This interpretation becomes apparent, if we set $\hat{c}_{i}^{(1)}=\hat{c}_{i}$
and $\hat{c}_{i}^{(2)}=\hat{c}_{i}^{\dagger}$ in the non-Abelian
AAH model. It leads to a Hamiltonian that describes a 1D $p$-wave
superfluid in a quasi-periodic potential, 
\begin{equation}
{\cal H=}\sum_{i}\left[\left(t_{x}\hat{c}_{i+1}^{\dagger}\hat{c}_{i}+\Delta\hat{c}_{i+1}\hat{c}_{i}+\text{H.c.}\right)+t_{y}V_{i}\hat{c}_{i}^{\dagger}\hat{c}_{i}\right],\label{eq:pwaveHami}
\end{equation}
which shares the same energy spectrum as the AAH model. The Anderson
localization of the Hamiltonian Eq. (\ref{eq:pwaveHami}) has been
recently analyzed by Lang \textit{et al.} \cite{Lang2012a} and DeGottardi
\textit{et al.} \cite{DeGottardi2013}. It was shown that the system
becomes localized once $t_{y}>t_{x}+\Delta$.

\subsection{Duality}

Our non-Abelian AAH model has an interesting duality, which is very
useful to understand the phase diagram. In the $p$-wave Hamiltonian
Eq. (\ref{eq:pwaveHami}), if we make the following replacement for
the field operator, 
\begin{eqnarray}
\hat{c}_{i} & \rightarrow & -\hat{d}_{i}^{\dagger}\quad\textrm{if \ensuremath{i\textrm{ is odd}}}\label{eq:DualA}
\end{eqnarray}
and 
\begin{equation}
\hat{c}_{i}\rightarrow\hat{d}_{i}\quad\textrm{if }i\textrm{ is even,}\label{eq:DualB}
\end{equation}
and replace $\phi$ by $\phi+1/2$ in the quasi-periodic potential
$V_{i}$, Eq. (\ref{eq:pwaveHami}) keeps the same form, except that
$t_{x}$ and $\Delta$ are exchanged. Thus, the system is self-dual
if $t_{x}=\Delta$.

In the absence of the $p$-wave order parameter ($\Delta=0$), the
AAH model also has Aubry-André duality, which relates a wave-function
at $\lambda\equiv t_{y}/t_{x}=a$ to the one at $\lambda=1/a$ by
a Fourier transformation \cite{Aubry1980}. From the phase diagram
Fig. \ref{fig1}, a similar duality seems to exist if $\Delta\neq0$.
Unfortunately, we are not able to find a simple transformation to
relate the parameters $t_{x}$ and $t_{y}$ for a given $\Delta$,
or the parameters $\Delta$ and $t_{y}$ for a given $t_{x}$.

\section{Inverse participation ratio and phase diagram}

\begin{figure}
\begin{centering}
\includegraphics[clip,width=0.48\textwidth]{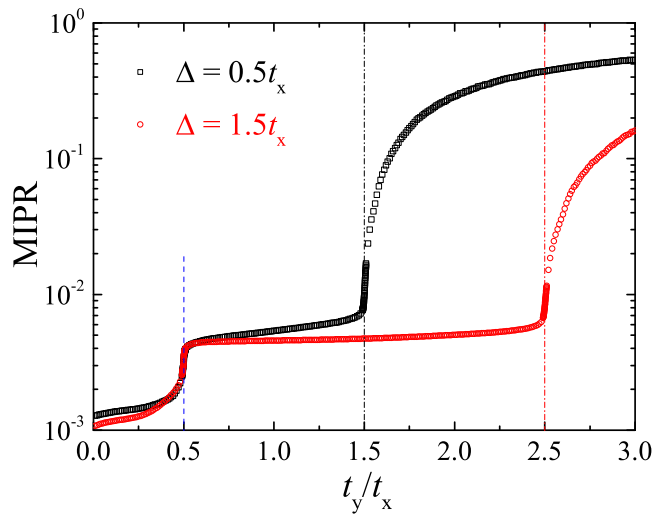} 
\par\end{centering}

\protect\caption{(Color online) Mean inverse participation ratio (MIPR) as a function
of the quasi-disorder strength at two $p$-wave superfluid order parameter
$\Delta=0.5t_{x}$ (black squares) and $\Delta=1.5t_{x}$ (red circles).
The dashed line and dot-dashed lines show the sharp increase of MIPR
at phase boundaries. Here, we use $n=15$ and $F_{n}=987$.}

\label{fig2} 
\end{figure}

A useful quantity in characterizing phase transitions of quasi-periodic
systems is the inverse participation ratio (IPR). For a given normalized
wave-function $\psi$, it is given by $\sum_{i}\psi_{i}^{4}=\sum_{i}[u_{i}^{4}+v_{i}^{4}]$,
which measures the inverse of the number of lattice sites being occupied
by particles. As we shall see, the non-Abelian AAH model considered
in this work has a $pure$ energy spectrum so that extended, critical
and localized wave functions do not coexist (i.e., there are no mobility
edges). Therefore, at the $n$-th rational approximation it is convenient
to define a mean inverse participation ratio (MIPR), 
\begin{equation}
\textrm{MIPR}=\frac{1}{2F_{n}}\sum_{i_{E}=0}^{2F_{n}-1}\sum_{i=0}^{F_{n}-1}\left[u_{i,i_{E}}^{4}+v_{i,i_{E}}^{4}\right],
\end{equation}
where $i_{E}$ is the index of energy levels. For extended states,
MIPR scales like $F_{n}^{-1}$; while for localized states, MIPR tends
to a finite value $O(1)$. For critical states, MIPR behaves like
$F_{n}^{-\alpha}$, where $0<\alpha<1$ depends on the multifractal
structure of wave-functions. We use MIPR to determine the phase boundaries
separating the extended, critical and localized phases, which are
identified by the turning points of MIPR as a function of $t_{y}/t_{x}$.

\begin{figure}
\begin{centering}
\includegraphics[clip,width=0.48\textwidth]{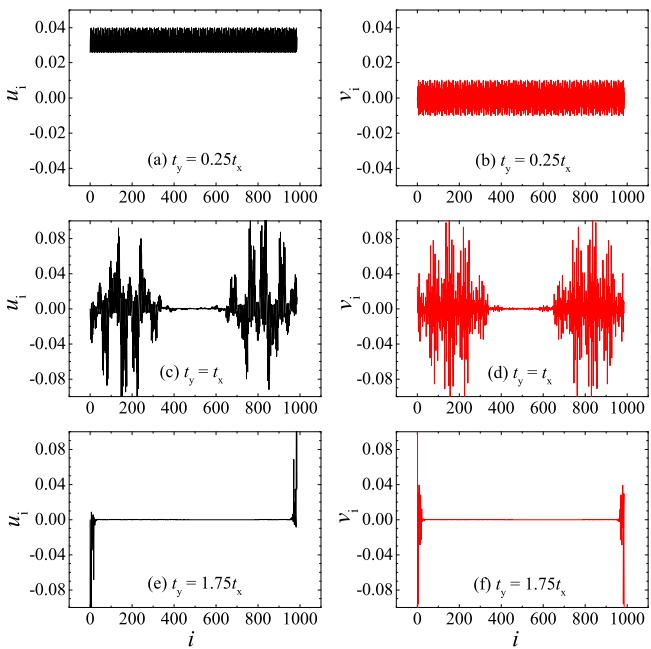} 
\par\end{centering}

\protect\caption{(Color online) The Bogoliubov quasi-particle wave-functions of the
lowest edge mode, $u_{i}$ (left panel) and $v_{i}$ (right panel),
at the $p$-wave superfluid order parameter $\Delta=0.5t_{x}$. From
top to bottom, the strength of the quasi-disorder $t_{y}$ increases
from $0.25t_{x}$ (a, b) to $t_{x}$ (c, d), and finally to $1.75t_{x}$
(e, f), corresponding to the cases of extended, criticial and localized
wave-functions, respectively. Here, we take $n=15$ and $F_{n}=987$.}

\label{fig3} 
\end{figure}

Figure \ref{fig2} reports the evolution of MIPR in a logarithmic
scale at two $p$-wave order parameters $\Delta=0.5t_{x}$ and $\Delta=1.5t_{x}$.
We have used $n=15$ and $F_{n}=987$. There are two turning points
of MIPR located respectively at $t_{y}=\left|t_{x}-\Delta\right|$
and $t_{x}+\Delta$, at which MIPR increases very rapidly. We have
checked that with increasing $n$, the slope of MIPR at these turning
points becomes sharper. Thus, we anticipate that finally there will
be a jump at the turning points in the scaling limit $n\rightarrow\infty$,
signaling a phase transition. The determination of the turning points
of MIPR at different $p$-wave order parameters leads to the proposed
phase diagram shown in Fig. \ref{fig1}.

We find four phases (I-IV) that are separated by three critical lines
$AB$, $AC$ and $CD$. Both phases I and IV have extended wave-functions
and are related to each other by the dual transformation given in
Eqs. (\ref{eq:DualA}) and (\ref{eq:DualB}). All the wave-functions
in the phase II are instead localized. It seems that these localized
wave-functions may also relate to the extended wave-functions in the
phase I (or IV) by a duality that is analogous to the Aubry-André
duality occurring at $\Delta=0$. Yet, such a duality transformation
is still to be determined. The three separating lines enclose a large
area in which all the wave-functions are critical.

In Fig. \ref{fig3}, we examine the representative ground-state wave-functions
(i.e., of the state at the edge of the energy spectrum) in different
phases with $\Delta=0.5t_{x}$. With increasing strength of the quasi-disorder
potential $t_{y}/t_{x}$, it is clear that the wave-function is extended
in the phase I (see a and b), critical in the phase III (c and d),
and localized in the phase II (e and f).

\section{Multifractal analysis of critical wave-functions}

To strengthen the proposed phase diagram in Fig. \ref{fig1}, we further
investigate the scaling behavior of wave-functions by using a multifractal
analysis \cite{Hiramoto1992,Hiramoto1989}. At the $n$-th level of
rational approximations, where the period of the lattice is $F_{n}$,
we analyze the probability measure at the lattice site $i$, $p_{i}=u_{i}^{2}+v_{i}^{2}$
($i=0,\cdots,F_{n}-1$), from a selected wave-function $\psi$, which
is normalized to unity $\sum_{i}p_{i}=1$. The scaling index $\alpha_{i}$
for $p_{i}$ is defined by 
\begin{equation}
p_{i}=F_{n}^{-\alpha_{i}}.
\end{equation}

The key observable to characterize the scaling behavior of the wave-function
is the singular spectrum $f_{n}(\alpha)$ defined by 
\begin{equation}
\Omega_{n}(\alpha)=F_{n}^{f_{n}(\alpha)},\label{eq:deffa}
\end{equation}
where $\Omega_{n}(\alpha)d\alpha$ is the number of lattice sites,
which have an index $\alpha_{i}$ distributed between $\alpha$ and
$\alpha+d\alpha$. The singular spectrum in the scaling limit $f(\alpha)$
can be calculated as $f(\alpha)=\lim_{n\rightarrow\infty}f_{n}(\alpha)$.
For extended wave-functions, all the lattice sites have a probability
measure $p_{i}\sim1/F_{n}$; thus $f(\alpha)$ is only defined at
$\alpha=1$ with $f(\alpha)=1$. For localized wave-functions, on
the other hand, $p_{i}$ is nonzero only on a finite number of lattice
sites. These sites have an index $\alpha=0$ and the remaining sites
with exponentially small probability measure have $\alpha=\infty$;
thus $f(\alpha)$ takes only two values: $f(\alpha=0)=0$ and $f(\alpha=\infty)=1$.
For critical wave-functions, the index $\alpha$ has a distribution
and hence the singular spectrum $f(\alpha)$ is a smooth function
defined on a finite interval {[}$\alpha_{\min}$, $\alpha_{\max}${]}.
Therefore, it is clear that, to identify extended, critical and localized
wave-function, we may simply examine the minimum value of the index
$\alpha$, which should take $\alpha_{\min}=1$ (extended), $1>\alpha_{\min}>0$
(critical) and $\alpha_{\min}=0$ (localized), respectively.

For the numerical calculation of $f(\alpha)$, we follow the work
by Hiramoto and Kohmoto \cite{Hiramoto1989} and define an entropy
function, 
\begin{equation}
\mathcal{S}_{n}\left(\alpha\right)=\frac{1}{n}\ln\Omega_{n}\left(\alpha\right),
\end{equation}
which is related to the singular spectrum as (cf. Eq. (\ref{eq:deffa}))
\begin{equation}
f_{n}\left(\alpha\right)=\frac{1}{\varepsilon}\mathcal{S}_{n}\left(\alpha\right),
\end{equation}
where $\varepsilon=\ln[(\sqrt{5}+1)/2]$. The entropy function can
be calculated by an analogous formalism to the usual statistical mechanics
\cite{Hiramoto1989}. First, one introduces a partition function,
\[
\mathcal{Z}_{n}\left(q\right)=\sum_{i=0}^{F_{n}-1}p_{i}^{q},
\]
and a free energy, 
\begin{equation}
\mathcal{G}_{n}\left(q\right)=\frac{1}{n}\ln\mathcal{Z}_{n}\left(q\right).
\end{equation}
The entropy function is then obtained through the Legendre transformation,
\[
\mathcal{S}_{n}\left(\alpha\right)=\mathcal{G}_{n}\left(q\right)+q\alpha\varepsilon
\]
and 
\begin{equation}
\alpha=-\frac{1}{\varepsilon}\frac{d\mathcal{G}_{n}\left(q\right)}{dq}.
\end{equation}

\begin{figure}
\begin{centering}
\includegraphics[clip,width=0.48\textwidth]{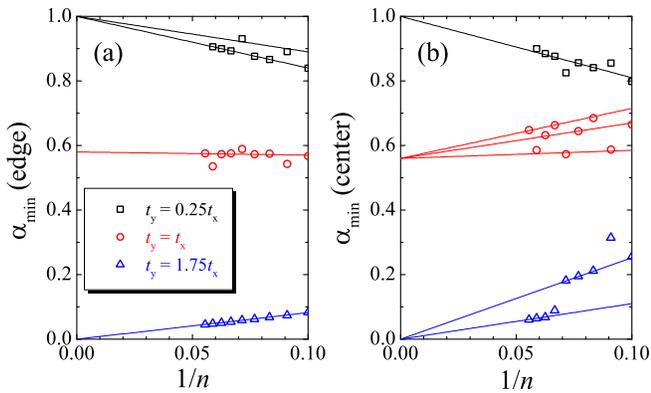} 
\par\end{centering}

\protect\caption{(Color online) Plots of $\alpha_{\min}$ vs $1/n$ for the wave-functions
of the lowest edge mode ($i_{E}=0$) (a) and of the center mode ($i_{E}=F_{n}$)
(b), at the $p$-wave superfluid order parameter $\Delta=0.5t_{x}$.
The black squares, red circles and blue triangles show the results
with increasing quasi-disorder strength $t_{y}$, as indicated in
the left panel (a).}

\label{fig4} 
\end{figure}

\begin{figure}
\begin{centering}
\includegraphics[clip,width=0.48\textwidth]{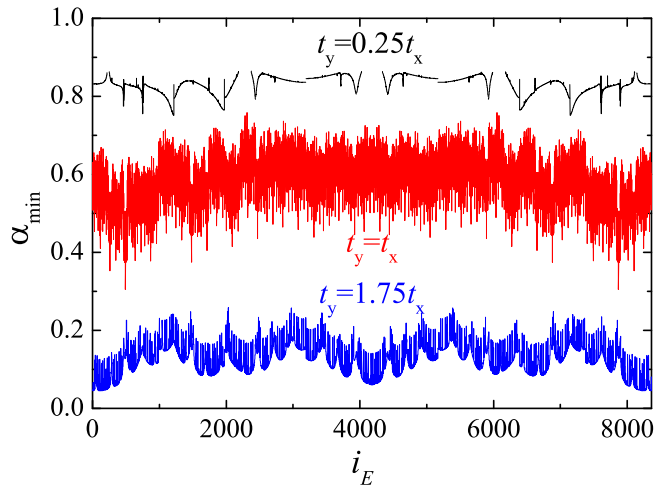} 
\par\end{centering}

\protect\caption{(Color online) Plots of $\alpha_{\min}$ for the wave-function of
each mode at $\Delta=0.5t_{x}$ and at three different quasi-disorder
strengths ($t_{y}/t_{x}=0.25$, $1$ and $1.75$ from top to bottom).
We have used $n=18$ and $F_{n}=4181$, so the index of energy mode
$i_{E}$ runs from $1$ to $2F_{n}=8362$. }

\label{fig5} 
\end{figure}

We have calculated $f_{n}(\alpha)=\mathscr{\mathcal{S}}_{n}(\alpha)/\varepsilon$
for finite Fibonacci indices $n$ and have tried to extrapolate them
to the scaling limit $1/n\rightarrow0$. In Fig. \ref{fig4}, we present
examples of determining $\alpha_{\min}$ for wave-functions in the
phases I, III and II at the $p$-wave order parameter $\Delta=0.5t_{x}$.
We consider two typical wave-functions, one at the edge of the energy
spectrum (denoted by an energy index $i_{E}=0$) and another at the
center of the spectrum ($i_{E}=F_{n}$). For both wave-functions,
$\alpha_{\min}$ extrapolates to $1$ if $t_{y}=0.25t_{x}$, to about
$0.58$ if $t_{y}=t_{x}$, and to $0$ if $t_{y}=1.75t_{x}$. In Fig.
\ref{fig5}, we examine the value of $\alpha_{\min}$ for all the
wave-functions at the $n=18\textrm{th}$ rational approximation. For
all three disorder strengths, $\alpha_{\min}$ varies smoothly with
increasing spectrum index $i_{E}$, suggesting that there is no mobility
edge in the energy spectrum. In this way, we confirm that all the
wave-functions are extended in the phase I, critical in the phase
III, and localized in the phase II.

\begin{table}
\protect\caption{Values of $\alpha_{\min}$ for the wave-functions at the edge and
the center of the energy spectrum. The mark $\times$ means that there
is no enough convergence in data for determining $\alpha_{\min}$.}

\centering{}%
\begin{tabular}{|c|c|c|c|c|}
\hline 
$\Delta/t_{x}$  & $t_{y}/t_{x}$  & position  & $\alpha_{\textrm{min}}$(edge)  & $\alpha_{\textrm{min}}$(center) \tabularnewline
\hline 
\hline 
0  & 1.0  & $A$  & 0.17  & 0.36\tabularnewline
\hline 
0.25  & 1.25  & $AB$  & $\times$  & 0.38\tabularnewline
\hline 
0.5  & 1.5  & $AB$  & $\times$  & 0.39\tabularnewline
\hline 
0.75  & 1.75  & $AB$  & $\times$  & 0.38\tabularnewline
\hline 
1.0  & 2.0  & $AB$  & $\times$  & 0.39\tabularnewline
\hline 
0.25  & 0.75  & $AC$  & 0.58  & 0.67\tabularnewline
\hline 
0.5  & 0.5  & $AC$  & 0.56  & 0.65\tabularnewline
\hline 
0.75  & 0.25  & $AC$  & 0.59  & 0.65\tabularnewline
\hline 
1.25  & 0.25  & $CD$  & 0.65  & 0.58\tabularnewline
\hline 
1.5  & 0.5  & $CD$  & 0.64  & 0.59\tabularnewline
\hline 
1.75  & 0.75  & $CD$  & 0.66  & 0.59\tabularnewline
\hline 
2.0  & 1.0  & $CD$  & 0.64  & 0.59\tabularnewline
\hline 
0.5  & 1.0  & III  & 0.58  & 0.56\tabularnewline
\hline 
1.0  & 0.5  & III  & 0.57  & 0.57\tabularnewline
\hline 
1.0  & 1.0  & III  & 0.57  & 0.56\tabularnewline
\hline 
1.0  & 1.5  & III  & 0.57  & 0.57\tabularnewline
\hline 
1.5  & 1.0  & III  & 0.57  & 0.58\tabularnewline
\hline 
1.5  & 1.5  & III  & 0.58  & 0.58\tabularnewline
\hline 
1.5  & 2.0  & III  & 0.57  & 0.58\tabularnewline
\hline 
2.0  & 1.5  & III  & 0.58  & 0.57\tabularnewline
\hline 
2.0  & 2.0  & III  & 0.57  & 0.57\tabularnewline
\hline 
\end{tabular}
\end{table}

We now focus on the wave-functions in the critical phase II and on
the three critical lines. Table I summarizes the values of $\alpha_{\min}$'s
for the edge and the center of the energy spectrum for various points
in the phase diagram, Fig. \ref{fig1}. At the duality point $A$
of the original AAH model ($\Delta=0,$ $t_{y}=t_{x}$), our results
of $\alpha_{\textrm{min}}\textrm{(edge)}\simeq0.17$ and $\alpha_{\textrm{min}}\textrm{(center)}\simeq0.36$
are consistent with the previous calculations by Hiramoto and Kohmoto
\cite{Hiramoto1989}. Quite generally, all the edge states on an individual
critical line, either $AB$, $AC$ or $CD$ (excluding the connecting
points $A$ and $C$), seem to have identical values of $\alpha_{\min}$.
This is also true for the center states, although their values of
$\alpha_{\min}$ may be different from those of the edge states. On
the other hand, all the states in the whole region III, no matter
at the edge or at the center of the spectrum, have identical values
of $\alpha_{\min}\simeq0.58$. Therefore, points on an individual
critical line (excluding the points $A$ and $C$) or in the whole
area III may belong to the same universality class. Further numerical
verification of this conjecture requires the comparison of the curve
$f(\alpha)$ on the whole interval {[}$\alpha_{\min}$, $\alpha_{\max}${]}.
Unfortunately, the convergence of our numerical estimates for $f(\alpha)$
at arbitrary $\alpha$ is too poor to reach a conclusive confirmation.

\section{Majorana edge modes}

\begin{figure}
\begin{centering}
\includegraphics[clip,width=0.48\textwidth]{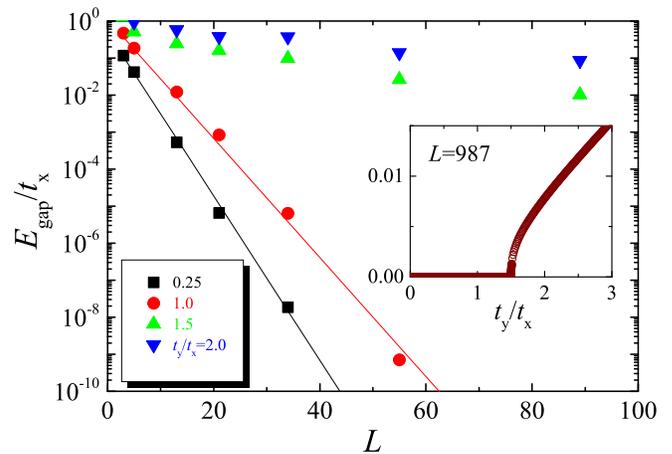} 
\par\end{centering}

\protect\caption{(Color online) Length dependence of the energy gap at $\Delta=0.5t_{x}$
and at different quasi-disorder strengths (as indicated). In the extended
and critical phases, the energy gap vanishes exponentially with increasing
system size, as suggested by the two exponential fitting lines. The
inset shows the evolution of the energy gap with increasing quasi-disorder
strength at $L=F_{n}=987$, where $n=15$. The energy gap increases
rapidly after entering the localized phase (II).}

\label{fig6} 
\end{figure}

In the absence of the quasi-periodic potential ($t_{y}=0$), the Schrödinger
equations (\ref{eq:SEa}) and (\ref{eq:SEb}) hosts zero-energy edge
mode at the two boundaries, as a result of its non-trivial topology
in the energy band structure. These zero-energy edge modes, also referred
to as Majorana fermions in the case of superfluidity, have been shown
to persist at nonzero $t_{y}$, until the localized phase is reached
\cite{DeGottardi2013,Cai2013}. We have calculated the energy of the
Majorana edge mode for a finite length system, $\epsilon_{M}$, by
using the open boundary condition, i.e., setting the $2\times2$ matrix
$C=0$ in Eq. (\ref{eq:matrixBdG}).

In Fig. \ref{fig6}, we report the energy gap $E_{gap}=2\epsilon_{M}$
as a function the system length $L=F_{n}$ at $\Delta=0.5t_{x}$ and
several quasi-disorder strengths. The length dependence of the edge-mode
energy is dramatically affected by the quasi-disorder strength. For
small disorder strengths in the extended phase I or the critical phase
III, the energy decreases exponentially as the length of the system
$L$ increases. On the other hand, on the separating critical line
$AB$ or in the localized phase II, the energy decreases much slower
with increasing length $L$. At a sufficiently large length $L=F_{15}=987$,
as shown in the inset of the figure, we find that the system opens
a nonzero energy gap only at the critical line $AB$, where $t_{y}>t_{x}+\Delta$,
in consistent with previous theoretical findings \cite{DeGottardi2013,Cai2013}.

\section{Conclusions}

In summary, we have proposed a generalization of the Aubry-André-Harper
model to the non-Abelian class with SU($N$) hopping matrices. The
localization and topological properties of the model is greatly affected
by such a generalization. We have performed a systematic investigation
of the simplest SU($2$) case with non-trivial $p$-wave superfluity
and have shown that its phase diagram becomes much richer. There is
a large window for the critical phase in the phase diagram, which
is separated from the extended and localized phases by three critical
lines. Our multifractal analysis of the critical wave-functions indicates
that points on an individual critical line or in the critical phase
may belong to the same universal class. Further issues - such as the
spectral statistics on the critical lines and in the critical phase
- are of interest and will be addressed elsewhere.

The proposed non-Abelian SU($2$) Aubry-André-Harper model might be
realized in cold-atom laboratories in the near future. In particular,
in view of recent numerous attempts for creating non-Abelian gauge
fields with ultracold atoms, various non-Abelian Aubry-André-Harper
models could be simulated. We anticipate even richer phase diagrams
with, for example, non-pure energy spectrum, in which the extended,
critical and localized states may coexist and be separated by some
mobility edges. Experimentally, it would be interesting to observe
mobility edges in one-dimensional quasi-periodic systems \cite{Ganeshan2015}. 
\begin{acknowledgments}
XJL and HH were supported by the ARC Discovery Projects (Grant Nos.
FT130100815, DP140103231, FT140100003, and DP140100637) and NFRP-China
(Grant No. 2011CB921502). GX was supported by the NSF of China (Grant
Nos. 11374266 and 11174253), the Zhejiang Provincial Natural Science
Foundation (Grant No. R6110175) and the Program for New Century Excellent
Talents in University.\end{acknowledgments}

\end{document}